\documentclass[10pt,conference]{IEEEtran}
\IEEEoverridecommandlockouts 
\usepackage{blkarray}                                   
\usepackage{graphicx}
\usepackage{svg}
\usepackage{amsmath}
\usepackage{amssymb}
\usepackage{amsfonts}
\usepackage{amsthm}
\usepackage[mathcal]{eucal}
\usepackage{mathrsfs}
\usepackage{booktabs}
\usepackage{siunitx}
\usepackage{enumerate}
\usepackage{multirow}
\usepackage[subrefformat=parens,farskip=0pt,justification=centering]{subfig}
\usepackage{color}
\usepackage{cite}                                          
\usepackage{comment}                                       
\usepackage{soul}                                          
\soulregister\cite7
\soulregister\ref7
\soulregister\pageref7
\usepackage{etoolbox}                                      
\usepackage{url}
\usepackage{nth}                                           
\usepackage{bm}                                            
\usepackage{courier}
\usepackage{balance}
\usepackage{threeparttable}
\usepackage{xcolor,colortbl}
\usepackage{footnote}
\usepackage{listings}
\usepackage{setspace}                                      
\usepackage[inline]{enumitem}
\usepackage{verbatim}
\usepackage[bookmarks=false]{hyperref}
\hypersetup{
    colorlinks = true,
    citecolor  = blue,
    linkcolor  = blue,
    urlcolor   = blue,
}
\usepackage{tikz}
\usetikzlibrary{patterns,snakes}
\usetikzlibrary{positioning,calc,fit,decorations.pathmorphing,shapes.geometric, shapes.gates.logic.US, calc}
\usetikzlibrary{arrows,arrows.meta,decorations.markings,shapes,shapes.arrows}
\usetikzlibrary{decorations,decorations.pathreplacing}
\usetikzlibrary{backgrounds}
\usepackage{filecontents}                                  
\usepackage{pgfplots}
\usepackage{pgfplotstable}
\usepgfplotslibrary{groupplots}
\usepackage{scalefnt}
\pgfplotsset{compat=newest}
\usepackage{caption}
\usepackage{pifont}                                        
\usepackage{cleveref}
\crefname{figure}{Fig.}{Figs.}
\Crefname{figure}{Fig.}{Figs.}
\crefname{table}{Table}{Tables}
\Crefname{table}{Table}{Tables}
\crefname{section}{Sec.}{Secs.}
\Crefname{section}{Sec.}{Secs.}
\crefname{subsection}{Sec.}{Secs.}
\Crefname{subsection}{Sec.}{Secs.}
\Crefformat{figure}{Fig.~#2#1#3}                           
\Crefname{subfigure}{Fig.}{Figs.}
\Crefname{figure}{Fig.}{Figs.}
\Crefformat{table}{TABLE~#2#1#3}                           
\usepackage[figuresright]{rotating}

\usepackage{algorithm}
\iftrue
\usepackage{algpseudocode}                                 
\algrenewcommand\textproc{\texttt}
\makeatletter
\let\OldStatex\Statex
\renewcommand{\Statex}[1][3]{%
  \setlength\@tempdima{\algorithmicindent}%
  \OldStatex\hskip\dimexpr#1\@tempdima\relax
}
\makeatother
\else
\usepackage{algorithmic}
\fi

\definecolor{CUHKorange}{RGB}{244,106,18} 
\definecolor{CUHKblue}{RGB}{0,111,190}    
\definecolor{CUHKgreen}{RGB}{0,127,128}   
\definecolor{CUHKred}{RGB}{228,46,36}     
\definecolor{CUHKyellow}{RGB}{198,148,34} 
\definecolor{CUHKdark}{RGB}{114,44,114}   
\definecolor{CUHKmiddle}{RGB}{144,44,144} 
\definecolor{CUHKlight}{RGB}{167,44,167} 
\definecolor{CUHKpurple}{RGB}{117,15,109}
\definecolor{CUHKgold}{RGB}{221,163,0}
\definecolor{CUHKribbon}{RGB}{244,223,176}
\definecolor{CUHKblack}{RGB}{34,24,21}

\renewcommand{\bf}[1]{\textbf{#1}}
\renewcommand{\tt}[1]{\texttt{#1}}

\newcommand{\minisection}[1]{\vspace{.01in}\noindent{\textbf{#1}}.}

\usepackage{tcolorbox}
\tcbuselibrary{skins,breakable}
    {\endtcolorbox}
    {\endtcolorbox}

\usepackage[top=0.80in,bottom=0.86in,left=0.62in,right=0.62in]{geometry}

\newcommand{\subparagraph}{}
\usepackage{titlesec}
\titlespacing*{\section}{0pt}{1.8ex plus .2ex minus .2ex}{0.4ex plus .2ex}
\titlespacing*{\subsection}{0pt}{1.0ex plus .2ex minus .2ex}{0.2ex plus .2ex}

\crefname{mytheorem}{Theorem}{Theorems}
\crefname{mylemma}{Lemma}{Lemmas}
\crefname{myclaim}{Claim}{Claims}
\crefname{myproperty}{Property}{Properties}
\crefname{mycorollary}{Corollary}{Corollaries}

\RequirePackage[normalem]{ulem} 
\RequirePackage{color}\definecolor{RED}{rgb}{1,0,0}\definecolor{BLUE}{rgb}{0,0,1} 

\usepackage{amsmath,amsfonts,bm}

\def\1{\bm{1}}

\def\ve{{\bm{e}}}

\def\vq{{\bm{q}}}
\def\vr{{\bm{r}}}

\def\vx{{\bm{x}}}

\def\mA{{\bm{A}}}

\def\mC{{\bm{C}}}

\def\mH{{\bm{H}}}

\def\vT{{\bm{T}}}

\def\mV{{\bm{V}}}

\DeclareMathAlphabet{\mathsfit}{\encodingdefault}{\sfdefault}{m}{sl}
\SetMathAlphabet{\mathsfit}{bold}{\encodingdefault}{\sfdefault}{bx}{n}

\graphicspath{{./figs/}{./}}

\usepackage[switch]{lineno}
\usepackage{tikz}
\usepackage{circuitikz}

\begin{document}

\title{
   Walk the Cycle: An Efficient Stochastic Solver for Periodic Thermal Analysis of 2.5D/3D ICs
}

\author{
  \IEEEauthorblockN{
    Shengnan Cai$^{1,*}$,
    Qibo Xue$^{1,*}$,
    Tianshu Hou$^{1}$,
    Tianxiang Zhu$^{2}$,\\
    Zixiao Wang$^{3,\dagger}$,
    Yibo Lin$^{2}$,
    Runsheng Wang$^{2}$,
    Bei Yu$^{1,\dagger}$
  }
  \IEEEauthorblockA{
    $^{1}$The Chinese University of Hong Kong, Hong Kong,
    $^{2}$Peking University, Beijing, China\\
    $^{3}$The Hong Kong University of Science and Technology (Guangzhou),
    Guangzhou, China
  }
}
\maketitle
\thispagestyle{plain}
\pagestyle{plain}

\begin{abstract}
Long-term periodic thermal analysis is essential for designing
2.5D/3D ICs operating under periodic workloads.
Deterministic global solvers can be costly for repeated thermal
queries, while existing random-walk methods are not tailored
to long-term periodic problems or multiscale 3D IC structures,
limiting their efficiency and accuracy. We propose \emph{CycleWalker}, a periodic residual random-walk solver.
First, we formulate the periodic steady state (PSS) as an infinite-history backward walk, which needs no initial temperature or warm-up but charges the full temperature field to every path.
Second, to cut this cost, we let the walk estimate only the residual of a periodic prior, and shorten long histories by Russian roulette while preserving unbiasedness.
Third, since the estimator constrains the prior only to be periodic and piecewise linear in time, we build it at negligible cost by a coarse-grid
Fourier solve followed by fine-grid smoothing.
Finally, we construct a thermal graph aware of through-silicon vias (TSVs) that bounds the walk's jump rate while retaining the anisotropic
conduction of TSV arrays.
Experiments on periodic 2.5D/3D IC benchmarks show that, at equal accuracy, \emph{CycleWalker} achieves up to $\mathbf{75.85\times}$ speedup over the transient random-walk solver.
\end{abstract}

\section{Introduction}
\label{sec:intro}

Thermal prediction is repeatedly required throughout IC design and optimization~\cite{vMa2021TAP,vWang2024ATPlace,vEris2018Chiplet}. In 2.5D/3D ICs, heat dissipation through complex structures, including slender through-silicon vias (TSVs), is a key design concern~\cite{vCong2005ThermalVia,vSridhar2010ThreeDICE}. These ICs, such as AI accelerators, often operate under periodic workloads~\cite{vLiu2006PeriodicThermal}, \textit{e.g.,} periodic inference batches~\cite{vKang2022DNNSAM} and task scheduling~\cite{vMoulik2021TARTS}. During sustained operation, their temperature settles into a periodic steady state (PSS)~\cite{vLiu2005PeriodicThermal}, whose variation over a cycle must be evaluated~\cite{vLiu2006PeriodicThermal}.

\begin{figure}[t]
\centering
\includegraphics[width=\columnwidth]{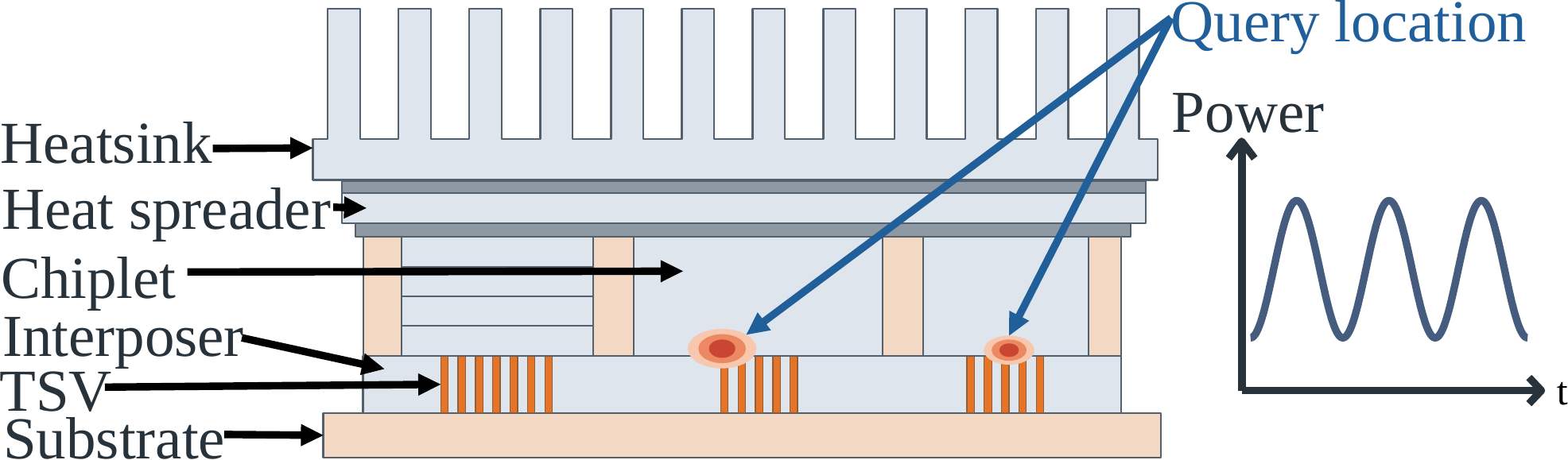}
\caption{Sparse queries $T(\vx,t)$ on a periodic 2.5D/3D IC stack.}
\label{fig:intro}
\end{figure}

Deterministic methods, such as the finite-volume method, are widely used for thermal analysis~\cite{vATSim3D2024}. Time-domain~\cite{vWang2002ThermalADI}, frequency-domain~\cite{vLiu2005PeriodicThermal}, and shooting methods~\cite{vTelichevesky1995ShootingPSS} can obtain the PSS, but their full-field computation can be costly at high spatial and temporal resolution~\cite{vLightSim2014,vChhabria2021ThermEDGe,vZhang2025FSAHeat,vLiu2023DeepOHeat,vYang2025MLTransient,vZhu2024FaStTherm}. In many design evaluations, however, only temperatures at a few locations and phases are needed, such as hotspot temperatures during high-power phases in \Cref{fig:intro}~\cite{vLiang2014HybridRW,ref_tpirw}. Much of the full-field computation is then spent on locations and phases that are not queried, motivating a solver that evaluates the PSS only where and when it is needed.

Random-walk methods~\cite{vLiang2014HybridRW,ref_tpirw} can reduce the cost of sparse thermal queries by estimating temperatures at selected locations without solving for the full field. Early floating and hybrid random-walk solvers~\cite{vLiang2014HybridRW} addressed steady-state thermal conduction, while subsequent Feynman--Kac-based solvers~\cite{vYang2022PIRW} incorporated Robin-boundary treatment and improved efficiency through precalculation~\cite{ref_ppirw} and GPU acceleration~\cite{vLuo2025GPU}. Prior-assisted random-walk methods further shorten paths by using an approximate temperature at the truncation point~\cite{ref_fastrw,ref_fasterrw}. These steady-state methods~\cite{ref_fastrw,ref_fasterrw}, however, do not capture the temporal evolution required for phase-dependent PSS analysis. Initial-value transient random-walk methods, including spacetime field solvers and time-backward walks on thermal networks~\cite{ref_tpirw}, incorporate time explicitly and use a prescribed initial temperature as the temporal endpoint~\cite{ref_tpirw}. In PSS analysis, no such initial temperature is given, so a backward walk must account for contributions from arbitrarily distant periods or rely on artificial initialization and warm-up. This motivates a random-walk solver that directly targets the PSS.

To address these challenges, we propose \emph{CycleWalker}, a random-walk
solver that directly targets PSS thermal analysis. We first formulate the
PSS as an infinite-history backward walk, which removes the need for an
initial temperature or warm-up, but charges the full temperature field to
every path. We therefore decompose the temperature into a periodic prior
and an error field, and let the walk estimate only the error, terminating
long histories by Russian roulette without bias. The estimator requires
the prior to be periodic and piecewise linear in time, but not accurate,
so we construct it at negligible cost by a coarse-grid Fourier solve
followed by fine-grid smoothing. Finally, since the walk runs on a thermal
graph, we construct one whose jump rate stays bounded across the
multiscale structures of 2.5D/3D ICs. Our main contributions are
summarized as follows:

\begin{itemize}[leftmargin=*]
\item \textbf{Periodic stochastic formulation.} To the best of our knowledge, this is the first random-walk formulation that directly targets long-term periodic thermal analysis.

\item \textbf{Residual Random Walk.} We decompose the temperature into a prior and an error field, and apply Random Walks to estimate the error. We combine unbiased Russian roulette with adaptive sampling to reduce path variance, avoid truncation bias, and set the path count.

\item \textbf{Periodicity-aware prior.} To meet the periodicity and piecewise linearity that the residual estimator requires, we construct a
low-cost prior via a coarse-grid Fourier solve followed by lightweight
fine-grid smoothing, so that the residual is sampled by the same periodic
walk with a closed-form time integral.

\item \textbf{TSV-aware graph construction.} We combine feature-aligned discretization with subgrid TSV homogenization. This bounds the TSV-induced jump rate of the walk while retaining the anisotropic conduction of TSV arrays in 2.5D/3D ICs.
\end{itemize}
\section{Preliminaries}
\label{sec:pre}

We begin with a review of discretized transient heat conduction model and show how its solution can be written as an expectation over Random Walks, which serves as the starting point of our periodic formulation.

\subsection{Transient Heat Conduction}
\label{sec:heat_conduction}

\begin{figure}[t]
\centering
\subfloat[Boundary partition\label{fig:boundary_conditions}]{%
  \includegraphics[width=.48\linewidth]{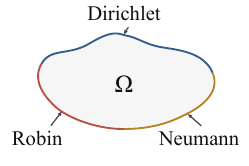}}\hfill
\subfloat[Walk on grid\label{fig:grid_walk}]{%
  \includegraphics[width=.48\linewidth]{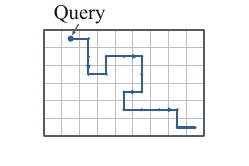}}
\caption{The domain $\Omega$ with various
boundaries, and a face-adjacent backward walk started from a query cell.}
\label{fig:prelim}
\end{figure}

Let $T(\vx,t)$ denote the temperature in the computational domain $\Omega$. It satisfies the transient heat conduction equation~\cite{vZhan2006Electrothermal}
\begin{equation}
c_v\frac{\partial T}{\partial t}-\nabla\cdot(k\nabla T)=q(\vx,t),
\label{eq:heat_equation}
\end{equation}
where $k$ is the thermal conductivity, $c_v$ is the volumetric heat capacity, and $q$ is the volumetric power density. The boundary $\partial\Omega$ is divided into three disjoint parts, $\partial\Omega=\Gamma_D\cup\Gamma_N\cup\Gamma_R$, on which different boundary conditions are imposed (\Cref{fig:boundary_conditions})~\cite{vYang2022PIRW}:
\begin{equation}
\begin{gathered}
T=T_D \ \text{on } \Gamma_D,
\qquad
k\frac{\partial T}{\partial n}=g_N \ \text{on } \Gamma_N,
\\
-k\frac{\partial T}{\partial n}=h(T-T_a) \ \text{on } \Gamma_R,
\end{gathered}
\label{eq:boundary_conditions}
\end{equation}
where $n$ is the outward unit normal. The three parts correspond to a prescribed boundary temperature $T_D$ (Dirichlet), a prescribed inward heat flux $g_N$ (Neumann, with $g_N=0$ being adiabatic), and convective cooling toward the ambient temperature $T_a$ with heat-transfer coefficient $h$ (Robin)~\cite{ref_tpirw}. In 2.5D/3D ICs, heat escapes through the top and bottom of the stack, so we lump the thermal resistance of the heat sink, TIM, and PCB into equivalent heat-transfer coefficients~\cite{vSridhar2014SemiAnalytical} and impose Robin boundaries there~\cite{vYuan2022PACT}; the lateral surfaces carry no cooling structure and are adiabatic. We denote by $B_i$ the conductance from cell $i$ to the external temperature through its boundary face. These Robin faces are the only absorption in the problem, so it is $B_i>0$ there that makes the path weight decay and the infinite history of \Cref{sec:periodic_formulation} converge at all.

A finite-volume discretization~\cite{vATSim3D2024} gives the heat balance of each cell $i$: the change of stored energy equals the heat flowing in from neighboring cells $j$, minus the heat lost through the boundary, plus the heat generated inside the cell,
\begin{equation}
C_i\frac{dT_i}{dt}=\sum_{j\in\mathcal N_i} G_{ij}(T_j-T_i)-B_i\bigl(T_i-T_{a,i}\bigr)+q_i,
\label{eq:cell_balance}
\end{equation}
where $\mathcal N_i$ is the set of cells sharing a face with cell $i$, $T_i$ is the temperature of cell $i$, $C_i=\int_{V_i}c_v\,dV$ is the heat capacity of cell $i$ with volume $V_i$, and $G_{ij}$ is the thermal conductance between cells $i$ and $j$. For a cell on $\Gamma_R$ or $\Gamma_D$,  $T_{a,i}$ is the external temperature on that face. The term $q_i$ is the power generated in the cell plus the Neumann flux through its faces. 
Collecting the unknown $T_i$ on the left and the known terms on the right yields~\cite{vLan2024TREND}
\begin{equation}
\mC\dot{\vT}(t)+\mA\vT(t)=\vq(t),
\label{eq:discrete_system}
\end{equation}
where $\mC$ is the diagonal capacitance matrix, $\mA$ collects inter-cell conduction and boundary dissipation, $\vq$ collects all known heat inputs, and $\vT(t)$ stacks the cell temperatures $T_i(t)$. \Cref{eq:discrete_system} is the system
on which the rest of this work operates. Our goal is to solve it for $\vT(t)$.

\subsection{Transient Random-Walk Representation}
\label{sec:transient_fk}

\begin{figure}[t]
\centering
\subfloat[Finite history\label{fig:finite_history}]{%
  \includegraphics[width=.48\linewidth]{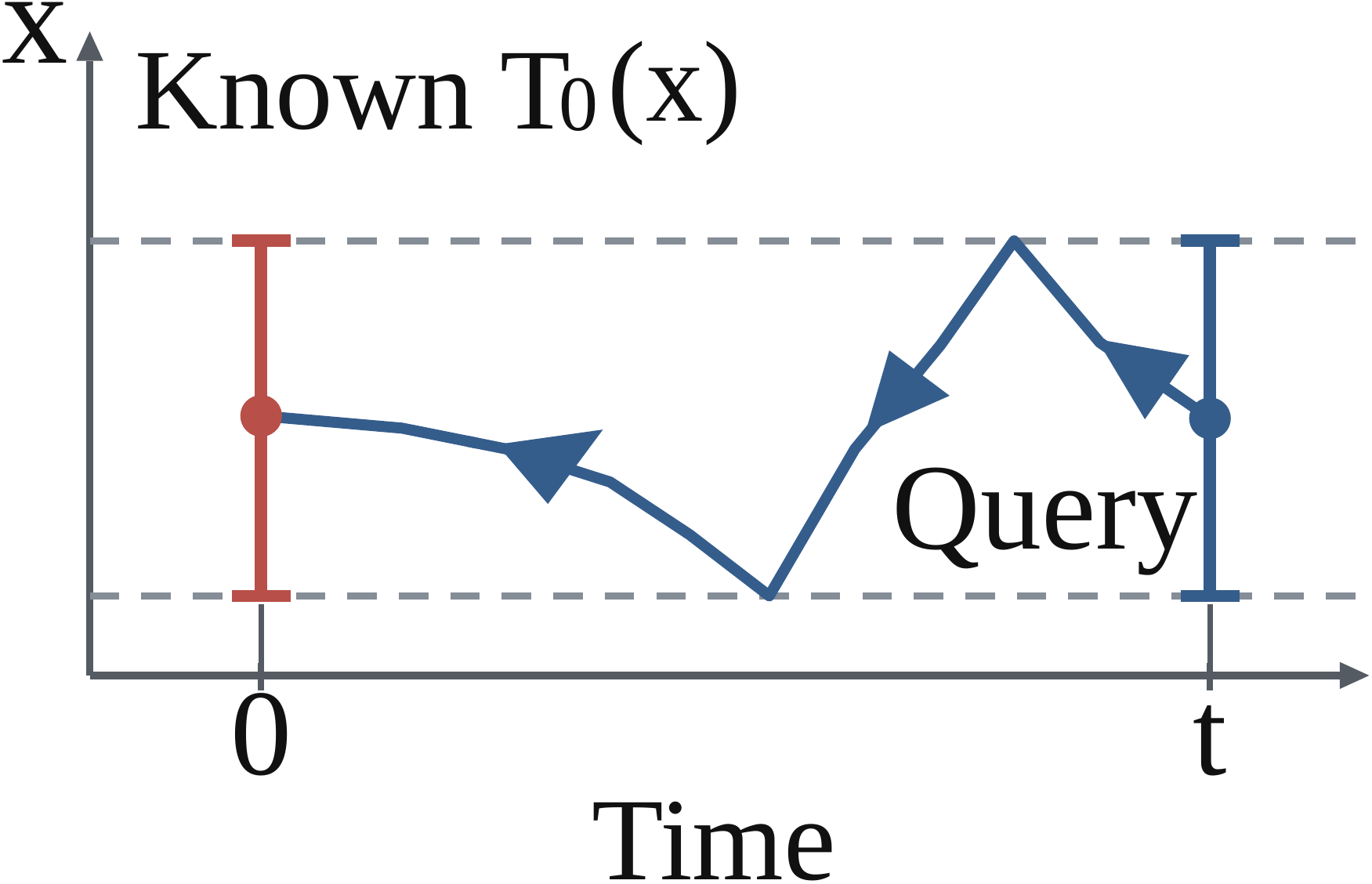}}\hfill
\subfloat[Periodic steady state\label{fig:periodic_history}]{%
  \includegraphics[width=.48\linewidth]{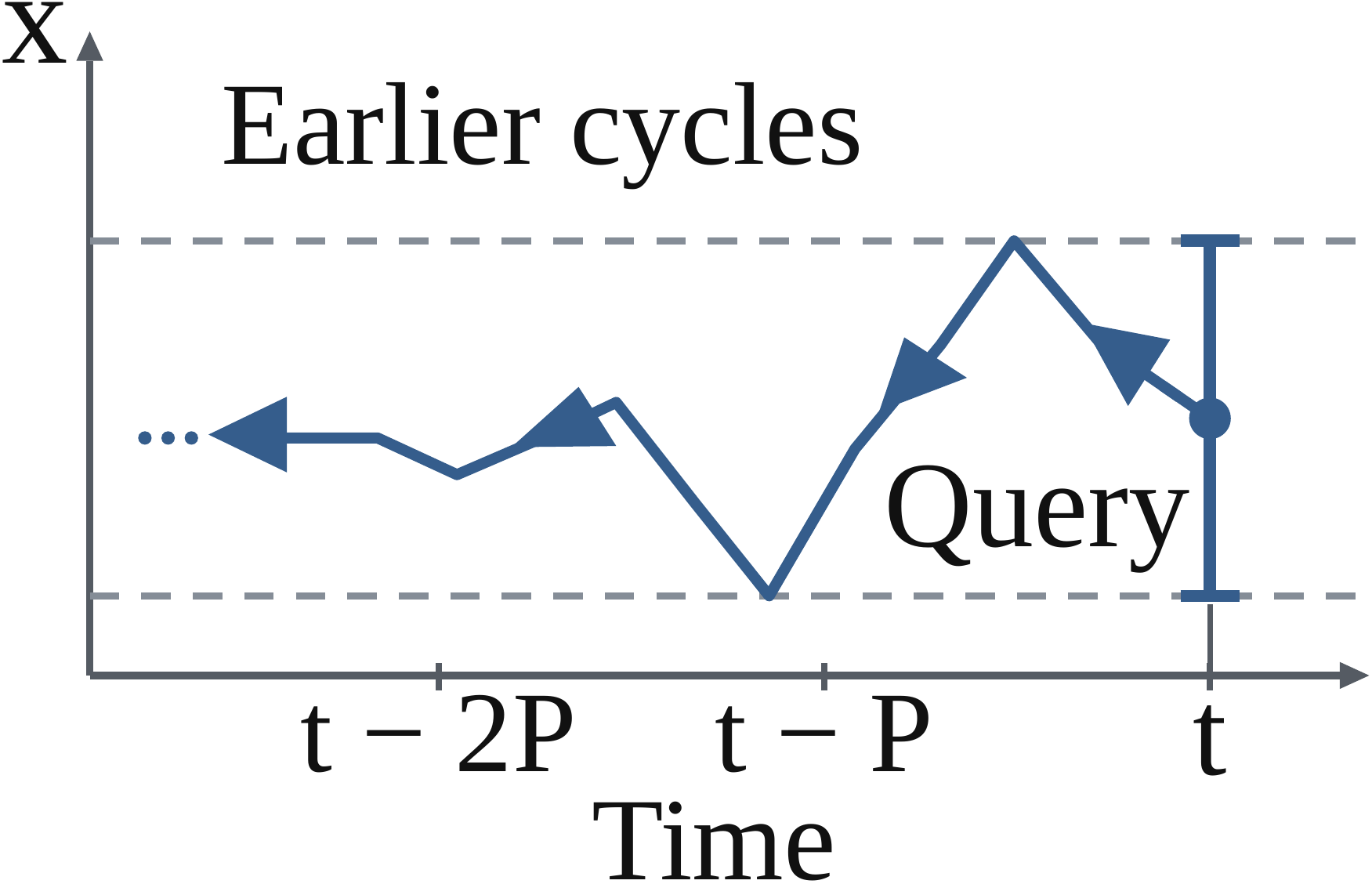}}
\caption{Backward walks in 1-D: finite history vs.\ endless PSS.}
\label{fig:pss_rw}
\end{figure}

\minisection{Walk on the Thermal Graph}
Random-walk methods~\cite{vLiang2014HybridRW,ref_tpirw} express the temperature as an expectation over stochastic paths. We adopt the walk-on-grid (WOG) formulation of transient random-walk solvers~\cite{vDong2025LCTMwalk}, in which a walker moves between neighboring cells of the thermal graph. By \Cref{eq:cell_balance}, the conduction term defines a continuous-time Markov chain, where a walker in cell $i$ jumps to neighbor $j$ at rate $G_{ij}/C_i$,
so the total jump rate $\nu_i$ out of the cell is
\begin{equation}
\nu_i=\frac{\sum_j G_{ij}}{C_i}.
\label{eq:jump_rate}
\end{equation}
The holding time in cell $i$ is exponentially distributed with mean $1/\nu_i$, so a cell with a large jump rate advances the walk by only a short backward time per step. Let $X_s$ denote the cell occupied at backward time $s$. The boundary term acts as absorption into the ambient at rate $B_i/C_i$. Instead of sampling this event, each path carries the survival weight
\begin{equation}
W_s=\exp\Big(-\int_0^s\frac{B(X_u)}{C(X_u)}\,du\Big),
\label{eq:fk_weight}
\end{equation}
which has the same expectation but lower variance~\cite{ref_tpirw}. Since $B_i=0$ in interior cells, $W_s$ decays only when the path visits boundary cells.
We use WOG rather than walk-on-spheres (WOS) throughout. A WOG step is face-adjacent, so the walk moves in the same way regardless of the material distribution (\Cref{fig:grid_walk}), and Robin boundaries enter exactly through $B_i$ without a tuned boundary layer~\cite{vYang2022PIRW,vDong2025LCTMwalk}. In the densely stacked thin layers of 2.5D/3D ICs, a WOS sphere cannot cross material interfaces anyway, so its radius would be capped by the layer thickness.

\minisection{Feynman--Kac Representation}
Walking backward from the query time $t$, the Feynman--Kac formula~\cite{ref_tpirw} gives
\begin{equation}
T_i(t)=\mathbb E_i\left[\int_0^t W_s\,\frac{q(X_s,\,t-s)}{C(X_s)}\,ds+W_t\,T_0(X_t)\right],
\label{eq:transient_feynman_kac}
\end{equation}
where $\mathbb E_i$ denotes the expectation over paths starting from cell $i$, $q(i,t)$ is the $i$-th entry of $\vq(t)$, and $T_0$ is the prescribed initial temperature. The path terminates at $t=0$, where the temperature is known (\Cref{fig:finite_history})~\cite{ref_tpirw}. Under long-term periodic operation, this endpoint no longer exists (\Cref{fig:periodic_history}). The walk therefore has no natural termination, which makes the periodic case substantially harder than the finite-history one, and it is precisely this case that we target in this work.
\section{Method}
\label{sec:method}
\emph{CycleWalker} solves a periodic thermal query in four stages. It first
writes the PSS temperature as an infinite-history backward walk
(\Cref{sec:periodic_formulation}). Since every path now carries the full
temperature field, the variance is very large, so many paths are needed. The walk is then built to estimate only the residual left by a
periodic prior, with unbiased early termination
(\Cref{sec:residual_rw}). With residual Random Walk, the prior does not need to be accurate, so it can be built cheaply in
the frequency domain and refined on the fine graph
(\Cref{sec:periodic_prior}). Finally, the walk runs on a thermal graph
whose jump rate stays bounded across the multiscale structures of 2.5D/3D
ICs (\Cref{sec:ic_adaptation}).

\subsection{Infinite-History Periodic Random-Walk Representation}
\label{sec:periodic_formulation}

For a workload with period $P$, $\vq(t+P)=\vq(t)$, we seek
the PSS satisfying $\vT(t+P)=\vT(t)$. The backward walk in
\Cref{eq:transient_feynman_kac} has to stop at a time where the
temperature is already known, but the PSS names no such time. The walk
in PSS must run back over several periods. So we push the starting time infinitely far into the past. The weight
$W_s$ decays as the walk accumulates conduction losses along the path,
so the contribution of the starting field vanishes and drops out of the
expectation:
\begin{equation}
T_i(t)=\mathbb E_i\left[\int_0^\infty W_s\,\frac{q\big(X_s,\,(t-s)\bmod P\big)}{C(X_s)}\,ds\right].
\label{eq:periodic_feynman_kac}
\end{equation} \Cref{eq:periodic_feynman_kac} is exact and needs no warm-up,
but it charges the full cost of the PSS to every path.

\subsection{Residual Random Walk}
\label{sec:residual_rw}

\begin{figure}[t]
\centering
\subfloat[$\vT$\label{fig:residual_T}]{%
    \includegraphics[width=0.32\columnwidth]{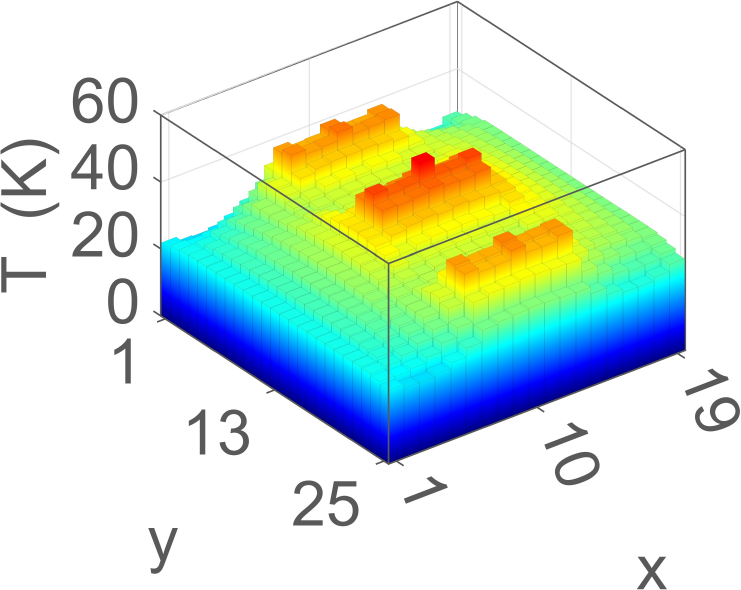}}
\hfill
\subfloat[$\vT_p$\label{fig:residual_Tp}]{%
    \includegraphics[width=0.32\columnwidth]{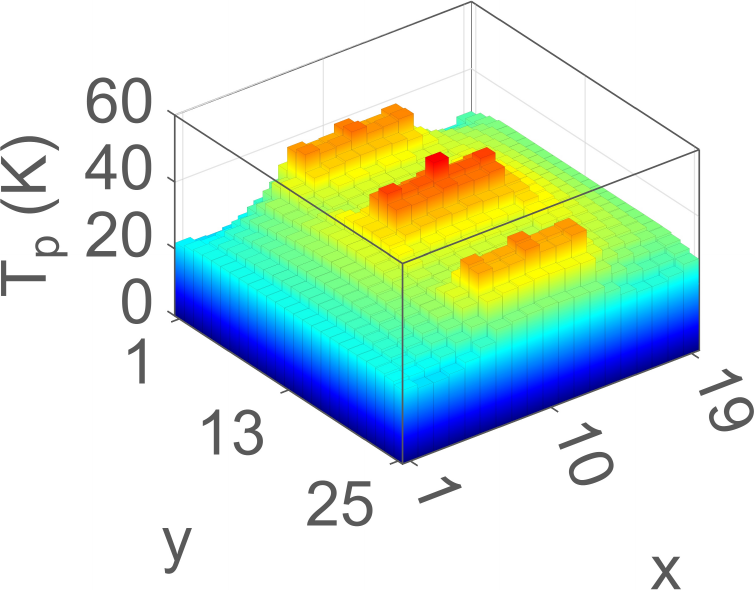}}
\hfill
\subfloat[$\ve=\vT-\vT_p$\label{fig:residual_e}]{%
    \includegraphics[width=0.32\columnwidth]{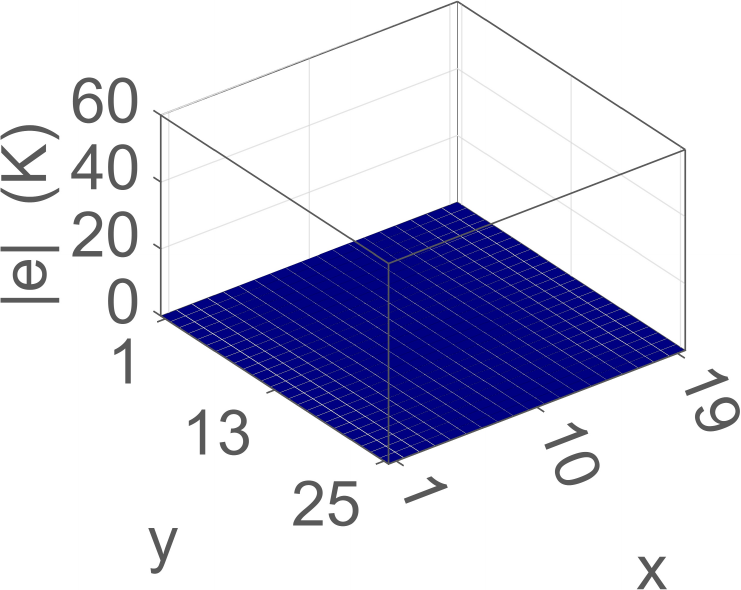}}
\caption{Residual decomposition: $\vT_p$ leaves $\ve$ small and smooth.}
\label{fig:residual}
\end{figure}

Three problems keep \Cref{eq:periodic_feynman_kac} from being
practical: each path reads the full power, so its variance is large
(\Cref{fig:residual_T}); the horizon is infinite, so any fixed
cutoff biases the estimate; and the variance differs a lot across queries, so a hand-picked path count either wastes work or silently misses the target accuracy. The
walk therefore estimates only the error of a cheap prior, Russian
roulette ends long paths without bias, and pilot paths set the budget.

\minisection{Residual formulation}
Let $\vT_p(t)$ be any $P$-periodic approximation of the PSS temperature, and let
$\ve(t)=\vT(t)-\vT_p(t)$ be its error. Substituting
$\vT=\vT_p+\ve$ into the discretized system of \Cref{eq:discrete_system},
whose capacitance matrix is $\mC$ and conduction matrix is
$\mA$, yields
\begin{equation}
\mC\dot{\ve}+\mA\ve=\vr,
\qquad
\vr=\vq-\mC\dot{\vT}_p-\mA\vT_p,
\label{eq:residual_system}
\end{equation}
where $\vr$ is the heat input left unexplained by the prior. 
The operator is unchanged, and $\vr$ inherits the $P$-periodicity
of $\vq$ and $\vT_p$, so $\ve$ is the PSS of \Cref{eq:residual_system}, and
\Cref{eq:periodic_feynman_kac} applies with $\vq$ replaced by
$\vr$:
\begin{equation}
e_i(t)=\mathbb E_i\left[\int_0^\infty W_s\,\frac{r\big(X_s,\,(t-s)\bmod P\big)}{C(X_s)}\,ds\right].
\label{eq:residual_estimator}
\end{equation}
Each path now reads the residual instead of the temperature. The
temperature is recovered by adding the prior back to the estimated
error. The estimators of $\vT$ and $\ve$ share the same form but not the same variance.
Paths now differ merely by the local error of $\vT_p$
(\Cref{fig:residual_e}), and the variance drops.

\minisection{Unbiased early termination}
We apply Russian roulette to the infinite-horizon integral in
\Cref{eq:residual_estimator}: if $W_s<\lambda$,
\begin{equation}
W_s\leftarrow
\begin{cases}
1, & \text{with probability } W_s,\\
0, & \text{otherwise (the path terminates),}
\end{cases}
\label{eq:russian_roulette}
\end{equation}
which leaves $\mathbb E[W_s]$ unchanged, so the estimator remains
unbiased while long paths are terminated early.

\minisection{Adaptive path budgeting}
A few pilot paths give the
largest sample variance $\sigma^2$ over all queries, and
\begin{equation}
N=\left\lceil\gamma\sigma^2/\varepsilon^2\right\rceil
\label{eq:required_paths}
\end{equation}
paths meet a target standard error $\varepsilon$, with $\gamma>1$
guarding against underestimation. This single budget is shared by all
queries, so none falls short of the target. The user therefore specifies
the target accuracy $\varepsilon$, and $N$ follows automatically.

The estimator places only two requirements on the prior. It must be
$P$-periodic, so that $\vr$ stays periodic in
\Cref{eq:residual_system}. It must be piecewise linear in time, so that
the residual integral in \Cref{eq:residual_estimator} is closed-form.
Accuracy is not required: the residual is computed from any given prior, so the estimator stays unbiased, and a cruder $\vT_p$ merely
costs more paths.

\subsection{Periodicity-Aware Prior Construction}
\label{sec:periodic_prior}

The prior must also be cheap, and neither standard route is. A transient solve reaches the PSS only after warm-up cycles, and a steady-state solve gives no phase dependence at all. We work in the frequency domain instead,
where a periodic source decouples the harmonics, so a coarse graph reaches the PSS cheaply.

\begin{figure}[t]
\centering
\subfloat[$\vT_c$: Coarse graph.\label{fig:prior_Tc}]{%
  \includegraphics[width=.5\linewidth,trim=0 0 0 -4]{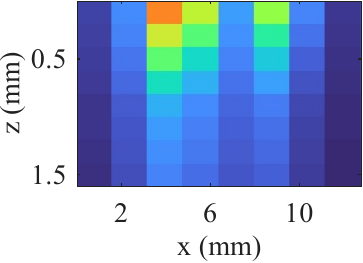}}\hfill
\subfloat[$\vT_{\mathrm{sp}}$: After sampling $\vT_c$.\label{fig:prior_Tsp}]{%
  \includegraphics[width=.5\linewidth,trim=-9 0 0 0]{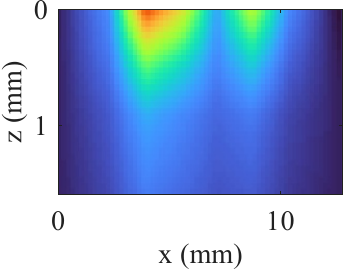}}\\[10pt]
\subfloat[$\vT_{\mathrm{sm}}$: After smoothing $\vT_{\mathrm{sp}}$.\label{fig:prior_Tsm}]{%
  \includegraphics[width=.5\linewidth,trim=-9 0 0 0]{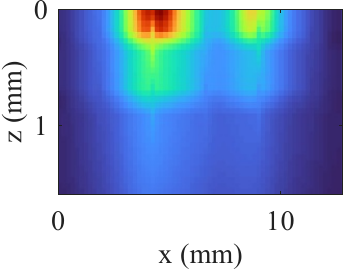}}\hfill
\subfloat[$\vT_p$: Slices along time.\label{fig:prior_Tp}]{%
  \includegraphics[width=.5\linewidth,trim=0 0 0 -8]{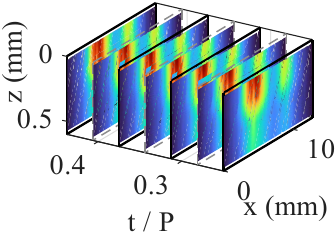}}
\caption{Prior construction on a slice of a die stack.}
\label{fig:prior_pipeline}
\end{figure}

A coarse graph alone is not enough, since it misses local detail near interfaces and TSVs. We therefore recover that detail by sampling and smoothing on the fine graph, and make the result continuous in time by interpolation.
\Cref{fig:prior_pipeline} illustrates the four stages. The prior is built as
\begin{equation}
\vT_c
\xrightarrow{\ \text{sample}\ }
\vT_{\mathrm{sp}}
\xrightarrow{\ \text{smooth}\ }
\vT_{\mathrm{sm}}
\xrightarrow{\ \text{interpolate}\ }
\vT_p,
\label{eq:prior_pipeline}
\end{equation}
where $\vT_c$ is the coarse-graph periodic solution, and $\vT_{\mathrm{sp}}$
and $\vT_{\mathrm{sm}}$ are fine-graph phase samples before and after
smoothing.

\minisection{Coarse graph}
The coarse graph merges each block of fine cells at a fixed ratio into one
coarse cell.
This shrinks the system by the cube of the ratio, and the loss of local
detail is acceptable because \Cref{sec:residual_rw} constrains the prior
only in form, not in accuracy.

On this graph we solve for the PSS directly in the frequency domain.
Since $\vq$ is $P$-periodic, expanding both $\vq$ and the PSS
$\vT_c$ in Fourier series with $\omega=2\pi/P$ turns the time derivative in
\Cref{eq:discrete_system} into a multiplication by $\mathrm j\ell\omega$.
The operator is linear and time-invariant, so the harmonics decouple and
each solves
\begin{equation}
(\mA_c+\mathrm j\ell\omega\,\mC_c)\,\vT_\ell=\vq_\ell,
\qquad \ell=0,\dots,L,
\label{eq:harmonic_solve}
\end{equation}
where $\mA_c$ and $\mC_c$ are the coarse-graph matrices, and
$\vT_\ell$ and $\vq_\ell$ are the $\ell$-th Fourier coefficients of
temperature and power. Every $\ell$ is an independent complex sparse solve
of coarse size, and summing the harmonics back gives $\vT_c(t)$. The periodicity is imposed by construction rather than reached by
time stepping, so no initial condition is needed and no warm-up cycle is
simulated: $L+1$ solves land on the PSS at once.
Higher harmonics are also cheap to discard, since $\|(\mA_c+\mathrm
j\ell\omega\mC_c)^{-1}\|$ decays with $\ell$, so the thermal response is
low-pass and a small $L$ suffices.

\minisection{Fine-graph refinement}
The coarse solution $\vT_c$ captures the global temperature envelope but
averages away both the spatial detail near hotspots and the sharp phase
variation of the periodic response, while a full fine-grid periodic
transient solve is prohibitively expensive. We exploit a structural
property of layered stacks: vertical coupling dominates lateral coupling,
so a smoother that treats only the vertical direction implicitly is both
cheap and effective, and because it decouples the columns, it maps
naturally onto the GPU.

We divide the period into $M$ intervals of length $\Delta t=P/M$ and
project $\vT_c$ at each phase onto the fine graph,
\begin{equation}
\vT_{\mathrm{sp}}(m)=\boldsymbol\Pi\,\vT_c(m\Delta t),
\quad m=0,\dots,M-1,
\label{eq:phase_sampling}
\end{equation}
where $\boldsymbol\Pi$ interpolates trilinearly from coarse to fine cell
centers. From here on the argument $m$ denotes the phase index rather
than a time, so that $\vT_{\mathrm{sp}}(m)$ is the fine-graph temperature
vector at phase $m\Delta t$, and indices are taken modulo $M$. We then split the fine-graph matrix $\mA=\mV+\mH$, where $\mV$ holds the diagonal and the vertical coupling and $\mH$ the lateral coupling, and starting from
$\vT_{\mathrm{sp}}(m)$ run a fixed number of periodic backward-Euler
sweeps
\begin{equation}
\Big(\mV+\frac{\mC}{\Delta t}\Big)\vT_{\mathrm{sm}}(m)
\leftarrow
\vq(m)+\frac{\mC}{\Delta t}\vT_{\mathrm{sm}}(m-1)-\mH\,\vT_{\mathrm{sm}}(m),
\label{eq:periodic_smoothing}
\end{equation}
where $\vq(m)$ averages $\vq$ over the interval ending at $m\Delta t$. Each sweep updates all phases from the previous iterate, so vertical
coupling is treated implicitly while lateral coupling and the preceding
phase are lagged.

Since $\mV$ couples only cells stacked in the same column, every column and
every phase is independent: one sweep reduces to a batch of $M$ times the
number of columns independent tridiagonal solves, executed as a single
GPU kernel. A fixed, user-specified number of sweeps therefore yields a
fine-grid periodic temperature field $\vT_{\mathrm{sm}}$ at deterministic
cost, which serves as the smooth deterministic component whose equation
residual drives the subsequent Random Walk. Comparing
\Cref{fig:prior_Tsp} with \Cref{fig:prior_Tsm} shows what the sweeps
recover: the hotspot sharpens and the layer interfaces and TSV columns
reappear, all lost in the coarse solve.

\minisection{Interpolation}
$\vT_p(t)$ linearly interpolates $\vT_{\mathrm{sm}}(m)$ between adjacent
phases (\Cref{fig:prior_Tp}). Then $\dot{\vT}_p$ and $\vq$ are both constant on each
interval, and the residual is linear in time there.

The prior is therefore $P$-periodic and piecewise linear in time, as
\Cref{sec:residual_rw} requires, and the residual integral needs no
further time discretization or sampling. It costs $L+1$ coarse solves and
a few column-parallel sweeps, and it absorbs the dominant profile, so
the walk samples a much smaller field.

\subsection{TSV-Aware Multiscale Graph Construction}

\newlength{\tsvh}
\setlength{\tsvh}{22mm}

\begin{figure}[t]
\centering
\subfloat[Via-resolving vs.\ feature-aligned.\label{fig:tsv_steps}]{%
  \includegraphics[height=\tsvh]{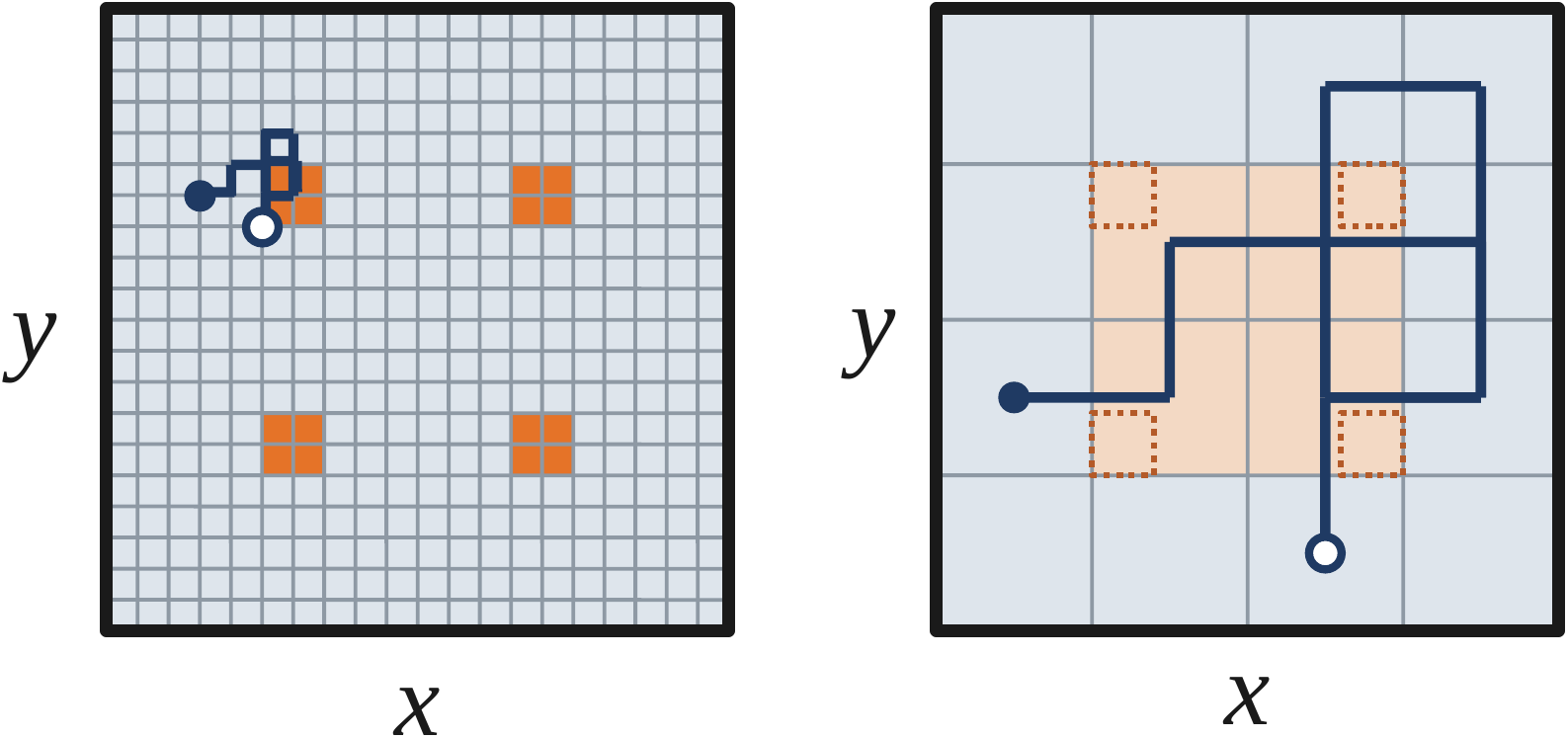}}\hfill
\subfloat[Walks in backward time.\label{fig:tsv_history}]{%
  \includegraphics[height=\tsvh]{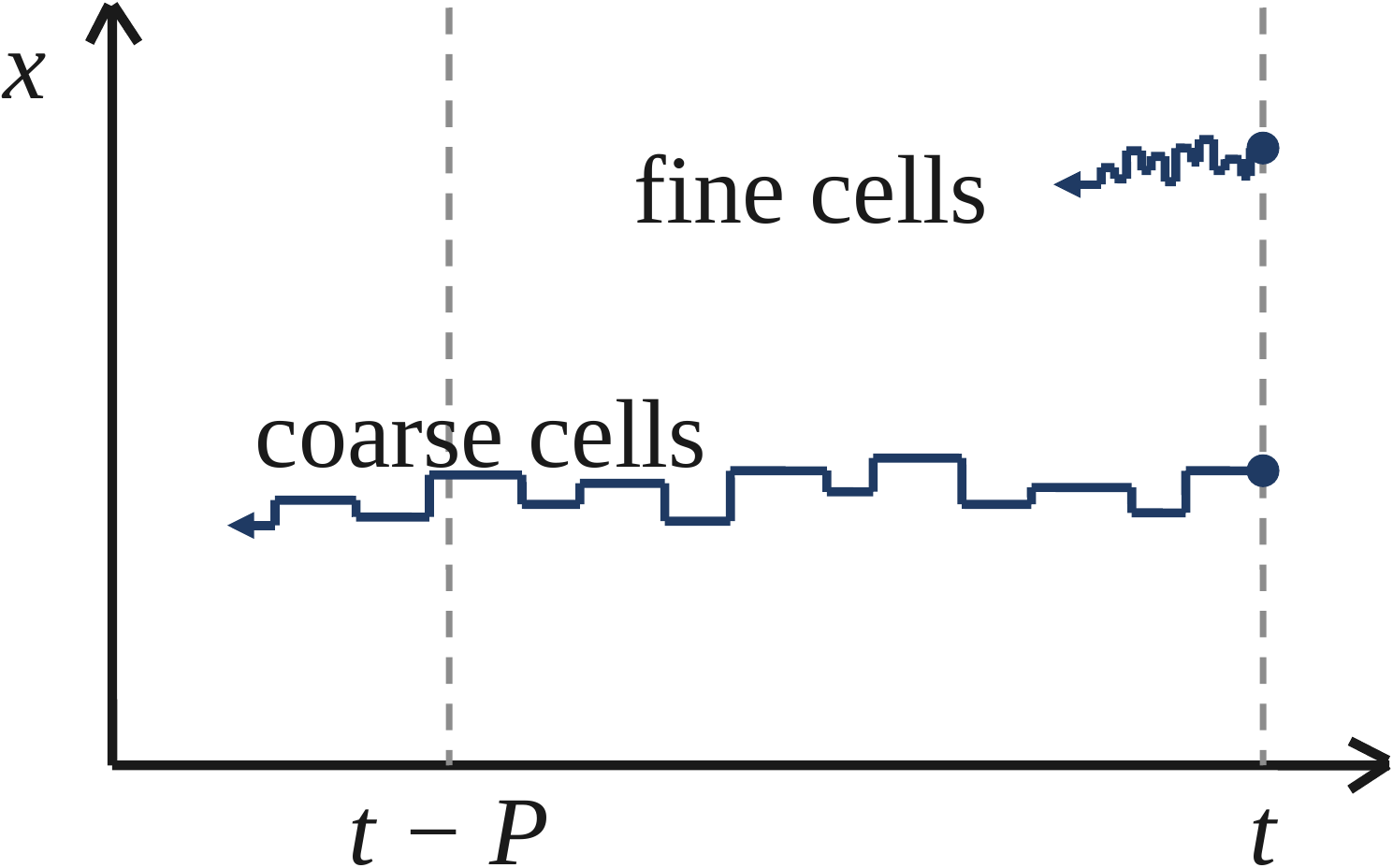}}
\caption{Feature-aligned grid bounds the jump rate; vias enter as material.}
\label{fig:tsv}
\end{figure}

\label{sec:ic_adaptation}

Resolving every micrometer-scale TSV with a conformal mesh would make
the walk prohibitively slow, because the jump rate is highly sensitive
to the cell thickness. By \Cref{eq:jump_rate}, a cell of thickness
$\Delta z$ has a conductance that scales as $k/\Delta z$ and a
capacitance that scales as $c_v\Delta z$, so
\begin{equation}
\frac{\sum_j G_{ij}}{C_i}\;\sim\;\frac{k}{c_v\,\Delta z^{2}}.
\label{eq:jump_rate_scaling}
\end{equation}
A conformal mesh would therefore raise the local jump rate by orders of
magnitude. Each step would advance only a tiny backward time, and
tracing back one period would require an enormous number of steps
(\Cref{fig:tsv_steps,fig:tsv_history}). The
vias cannot be dropped either: a via array conducts heat much better
along the vias than across them, and this directional behavior dominates
heat removal in stacked packages. The graph must bound the jump rate
yet keep this anisotropy.

\minisection{Feature-aligned discretization}
We use a nonuniform Cartesian grid that preserves material interfaces
and heat-source boundaries, and never refines to the via scale. The cell
size is then set by the grid resolution rather than by the smallest TSV,
and the jump rate is bounded accordingly.

\minisection{Subgrid TSV homogenization}
We keep the vias as a material property instead of a geometry. Let
$\chi$ be the area fraction of a cell's cross section occupied by vias,
and let $k_h$ and $k_c$ be the conductivities of the host material and
the copper fill. The effective lateral conductivity $\kappa$ and axial
conductivity $\eta$ are
\begin{equation}
\kappa=\Big(\frac{1-\chi}{k_h}+\frac{\chi}{k_c}\Big)^{-1},
\qquad
\eta=(1-\chi)\,k_h+\chi\,k_c,
\label{eq:tsv_mixing}
\end{equation}
and the volumetric heat capacity is mixed linearly. Both materials
are isotropic, but the cell is not: heat moves more easily along the vias than across them.

We compute the conductances $G$ from $\kappa$ and $\eta$ in the usual
way, so the walk needs no special rule for homogenized cells: the shaded
cells in \Cref{fig:tsv_steps} carry the vias without resolving them. This
layered structure is also what the vertical block smoothing of
\Cref{sec:periodic_prior} exploits.
\begin{figure*}[t]
\centering
\subfloat[3DIC-FastCycle\label{fig:mae_vs_N_fastcycle}]{%
    \includegraphics[height=0.1773\textwidth]{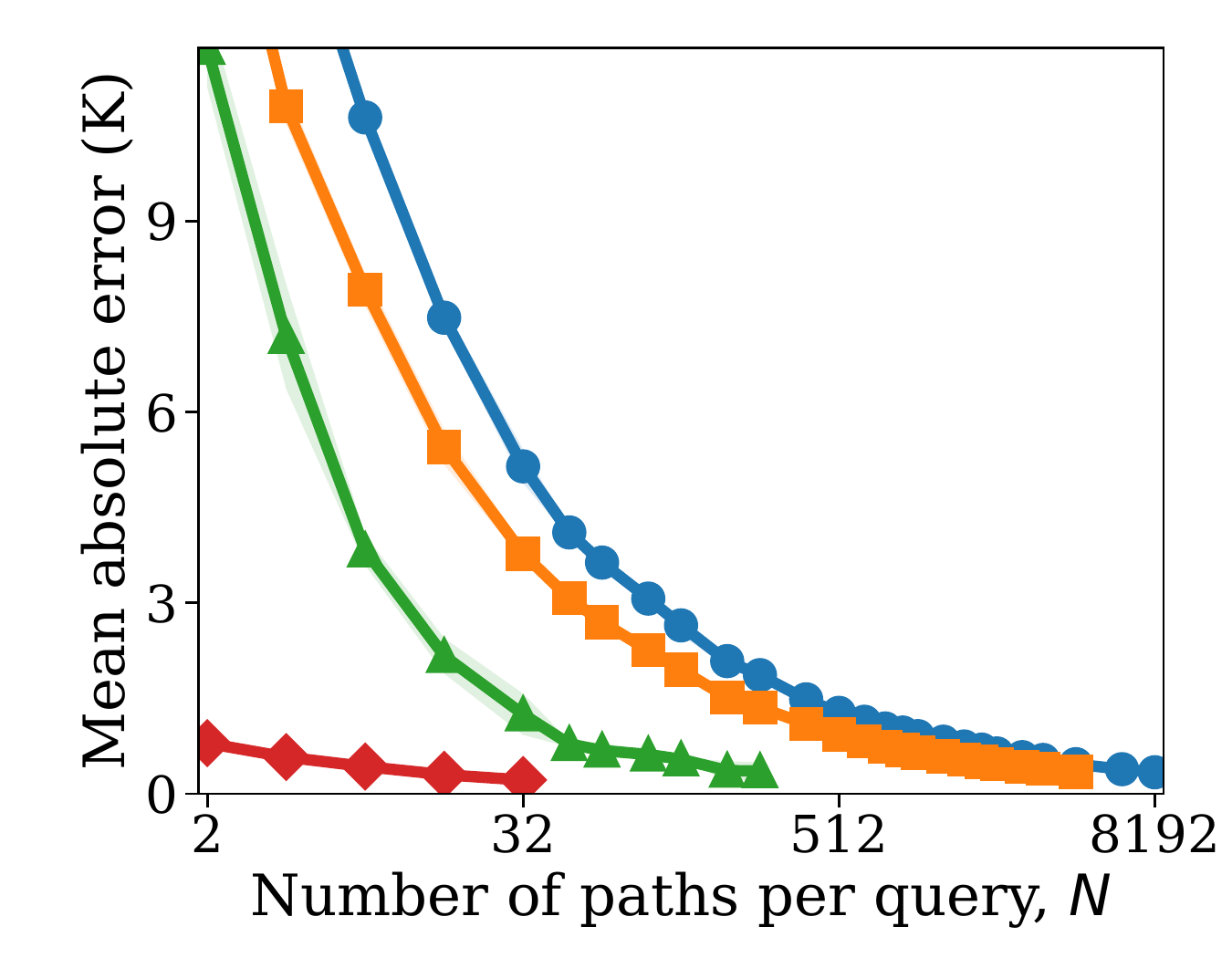}}
\hspace{0.01\textwidth}
\subfloat[3DIC-NonAligned\label{fig:mae_vs_N_nonaligned}]{%
    \includegraphics[height=0.1773\textwidth]{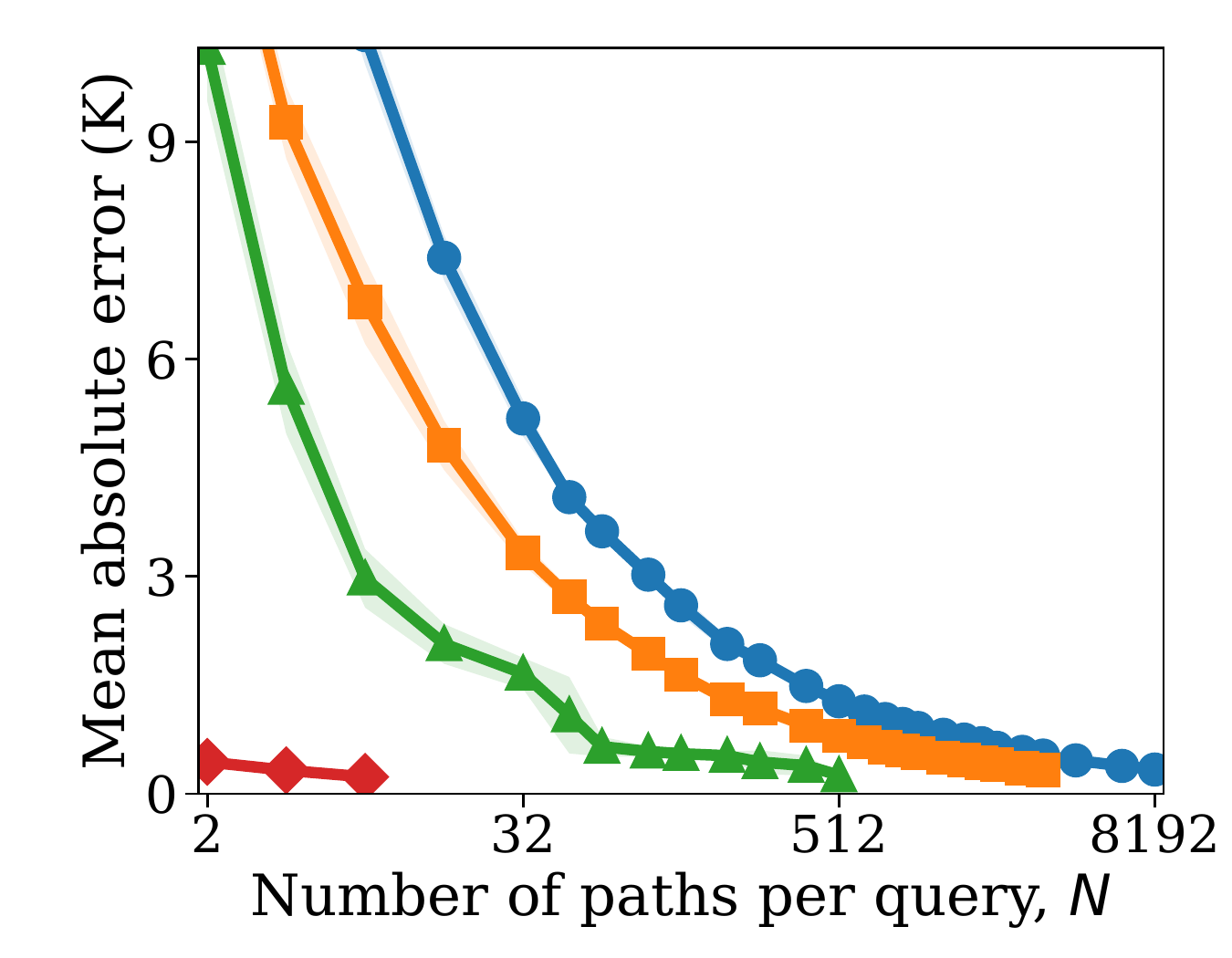}}
\hspace{0.01\textwidth}
\subfloat[GPU-HBM-Burst\label{fig:mae_vs_N_gpuhbm}]{%
    \includegraphics[height=0.1773\textwidth]{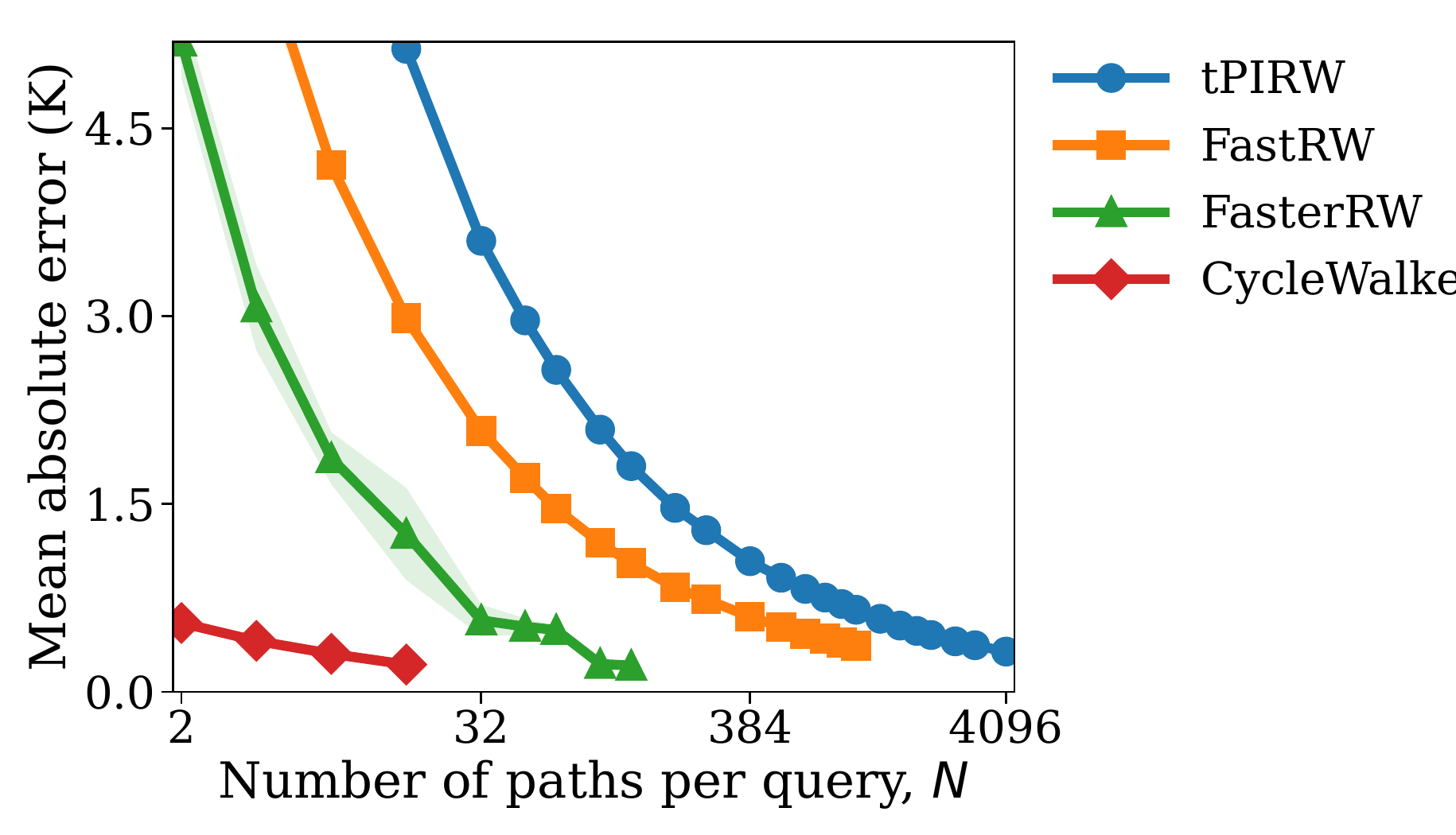}}
\caption{Error versus path budget on the three scenarios.}
\label{fig:mae_vs_N}
\end{figure*}

\section{Experiments}
\label{sec:exp}

We evaluate \emph{CycleWalker} on three periodic thermal scenarios: first against three stochastic baselines under identical accuracy constraints, then by a cumulative ablation of its estimator, and finally on the TSV-aware graph construction.

\subsection{Experimental Setup}
\label{subsec:exp_setup}

\begin{figure}[!t]
\centering
\subfloat[GPU-HBM-Burst package\label{fig:cases_gpuhbm}]{%
    \includegraphics[width=0.45\columnwidth]{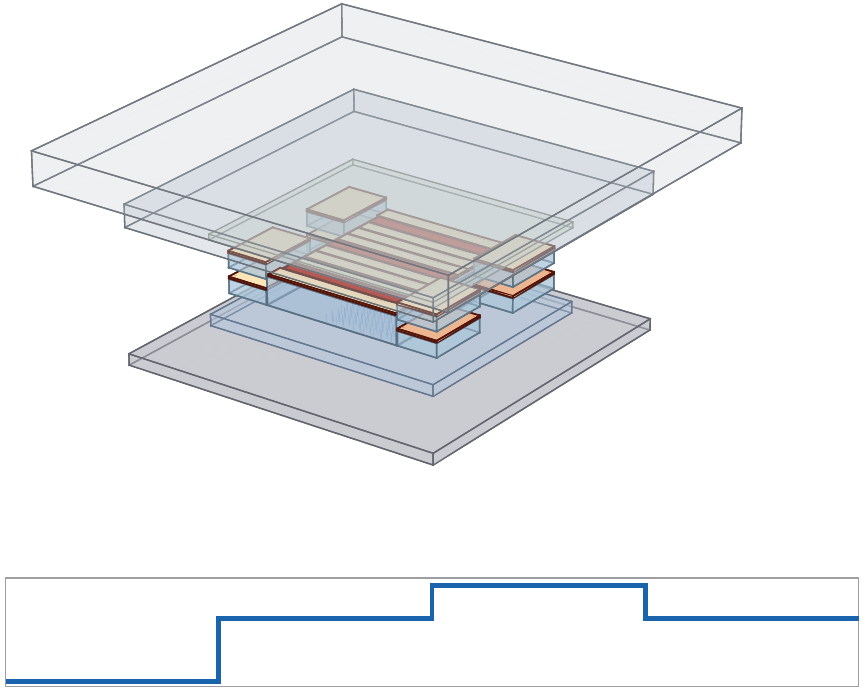}}
\subfloat[GPU-HBM-Burst floorplan\label{fig:cases_gpuhbm_fp}]{%
    \includegraphics[width=0.45\columnwidth]{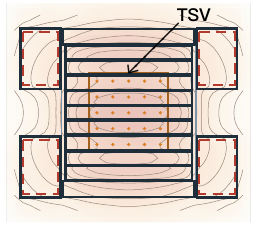}}
\\[2pt]
\subfloat[3DIC package\label{fig:cases_3dic}]{%
    \includegraphics[width=0.45\columnwidth]{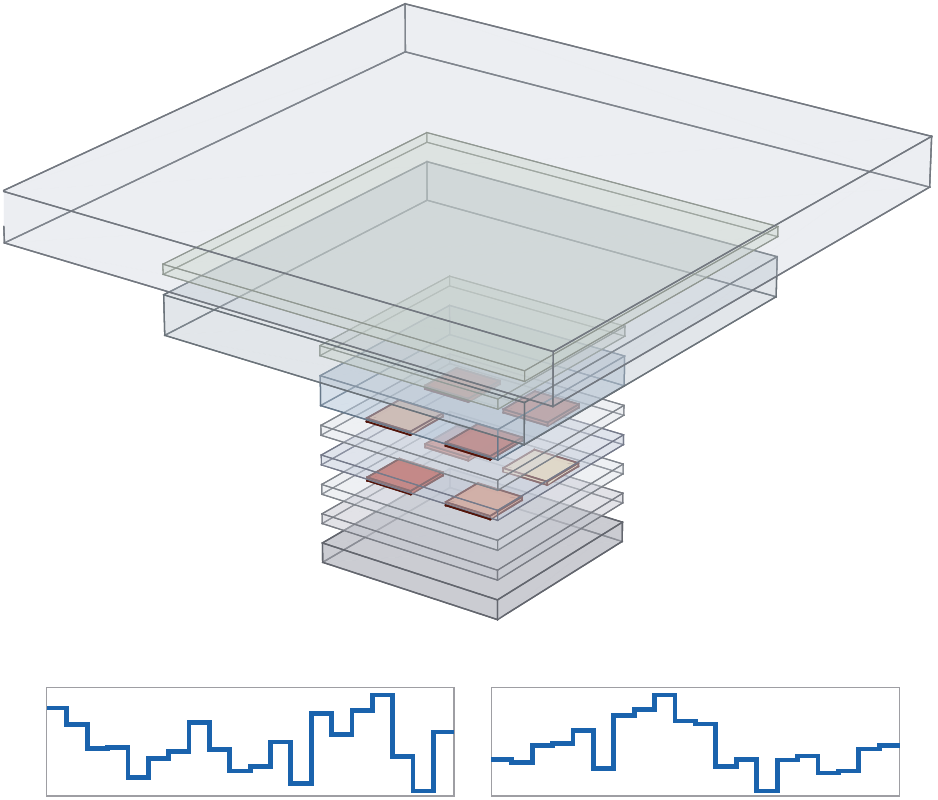}}
\subfloat[3DIC floorplan\label{fig:cases_3dic_fp}]{%
    \includegraphics[width=0.45\columnwidth]{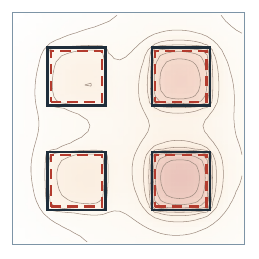}}
\caption{Benchmark cases. (a)(c) Exploded package views with the
one-period total power below each stack. (b)(d) Floorplans at the lower
active plane: unit outlines over the TSV array, shaded by
the reference steady-state temperature on that plane.}
\label{fig:cases}
\end{figure}

\minisection{Test Cases} \Cref{fig:cases} shows the two packages.
3DIC-NonAligned and 3DIC-FastCycle are the same ten-layer 3D stack
(35,600 cells) under \SI{0.2}{\second} and \SI{0.08}{\second} periods, the
former with query phases falling between switching instants.
GPU-HBM-Burst is a 2.5D TSV-interposer package (82,420 cells) with a
$31\times31$ TSV array and 19 power units on three active planes, driven
by a bursty workload. Power models are frozen at a reference temperature of \SI{300}{\kelvin}, so the problem is linear.

\minisection{Implementation}
Accuracy is measured against a 512-harmonic deterministic solution on the identical discretized operator, with relative residuals below $1.1\times10^{-13}$. All four methods share one CUDA implementation and one host (Intel Core i5-12500H, NVIDIA RTX 3050 Ti Laptop GPU); references are generated offline and excluded from all timings. \textbf{\emph{CycleWalker}} uses an 8-harmonic coarse prior, whole-path equation-residual estimation, and unbiased Russian roulette. FastRW and FasterRW use a cutoff of $0.5$.

\begin{table}[!t]
\centering
\caption{Iso-accuracy comparison at two targets.}
\label{tab:iso_accuracy}
\small
\setlength{\tabcolsep}{4pt}
\begin{tabular}{@{}llcrrr@{}}
\toprule
$\varepsilon$ (K) & Method & Err$\pm$Std (K) & $N$ & Time (s) & Speedup \\
\midrule
\rowcolor{gray!15}
\multicolumn{6}{@{}l}{\textit{3DIC-NonAligned}} \\
\multirow{4}{*}{0.2}
 & tPIRW    & $0.190\pm0.008$ & 24576 & 806.97 & 1.00 \\
 & FastRW   & $0.213\pm0.006$ & 8192  & 134.17 & 6.01 \\
 & FasterRW & $0.193\pm0.059$ & 1260  & 99.62  & 8.10 \\
 & \emph{CycleWalker} & $0.215\pm0.006$ & \bfseries 9 & \bfseries 17.09 & \bfseries 47.22 \\
\cmidrule(l){2-6}
\multirow{4}{*}{0.4}
 & tPIRW    & $0.380\pm0.002$ & 6144 & 199.69 & 1.00 \\
 & FastRW   & $0.419\pm0.011$ & 2048 & 41.32  & 4.83 \\
 & FasterRW & $0.355\pm0.141$ & 384 & 38.27 & 5.22 \\
 & \emph{CycleWalker} & $0.362\pm0.015$ & \bfseries 3 & \bfseries 14.71 & \bfseries 13.58 \\
\midrule
\rowcolor{gray!15}
\multicolumn{6}{@{}l}{\textit{3DIC-FastCycle}} \\
\multirow{4}{*}{0.2}
 & tPIRW    & $0.191\pm0.011$ & 24576 & 746.63 & 1.00 \\
 & FastRW   & $0.200\pm0.004$ & 12288 & 178.74 & 4.18 \\
 & FasterRW & $0.204\pm0.018$ & 1536  & 108.19 & 6.90 \\
 & \emph{CycleWalker} & $0.214\pm0.008$ & \bfseries 32 & \bfseries 17.52 & \bfseries 42.62 \\
\cmidrule(l){2-6}
\multirow{4}{*}{0.4}
 & tPIRW    & $0.381\pm0.009$ & 6144 & 182.56 & 1.00 \\
 & FastRW   & $0.397\pm0.013$ & 3072 & 54.42  & 3.35 \\
 & FasterRW & $0.414\pm0.075$ & 236  & 28.38  & 6.43 \\
 & \emph{CycleWalker} & $0.413\pm0.017$ & \bfseries 8 & \bfseries 17.80 & \bfseries 10.26 \\
\midrule
\rowcolor{gray!15}
\multicolumn{6}{@{}l}{\textit{GPU-HBM-Burst}} \\
\multirow{4}{*}{0.2}
 & tPIRW    & $0.186\pm0.004$ & 12288 & 3309.53 & 1.00 \\
 & FastRW   & $0.210\pm0.004$ & 3072  & 258.76 & 12.79 \\
 & FasterRW & $0.210\pm0.029$ & 128   & 116.50 & 28.41 \\
 & \emph{CycleWalker} & $0.218\pm0.003$ & \bfseries 16 & \bfseries 43.63 & \bfseries 75.85 \\
\cmidrule(l){2-6}
\multirow{4}{*}{0.4}
 & tPIRW    & $0.405\pm0.011$ & 2560 & 704.64 & 1.00 \\
 & FastRW   & $0.386\pm0.009$ & 896  & 128.60 & 5.48 \\
 & FasterRW & $0.438\pm0.089$ & 104 & 89.70 & 7.86 \\
 & \emph{CycleWalker} & $0.411\pm0.009$ & \bfseries 4 & \bfseries 56.09 & \bfseries 12.56 \\
\bottomrule
\end{tabular}
\end{table}

\minisection{Baselines}
\textbf{tPIRW}~\cite{ref_tpirw} is a periodic Feynman--Kac walk without a prior, so its truncation residual scales with the absolute temperature and forces a cutoff of $3\times10^{-4}$. \textbf{FastRW}~\cite{ref_fastrw} adds a periodic terminal prior, cross-query tail reuse, and Bayesian fusion. \textbf{FasterRW}~\cite{ref_fasterrw} further adds covariance estimation and periodic kriging refinement. 

\minisection{Metrics}
Accuracy and cost are reported as the mean absolute error (MAE) and the speedup, 
\begin{equation}
\mathrm{MAE}=\frac{1}{Q}\sum_{j=1}^{Q}\big|T_j-T_j^{\mathrm{ref}}\big|,
\qquad
\mathrm{Speedup}=\frac{T_{\mathrm{tPIRW}}}{T_{\mathrm{method}}}.
\label{eq:metrics}
\end{equation}
Speedup uses end-to-end solver time, since prior-assisted methods shift cost
into deterministic stages; graph construction and references are excluded.
Var.\ is the per-path variance of a query and $\mathrm{SE}=\sqrt{\mathrm{Var}/N}$
its standard error.

\subsection{Iso-Accuracy Comparison}
\label{subsec:exp_overall}
Budgets are selected on three calibration seeds for each nominal target
$\varepsilon$, and the selected configurations are then evaluated on five
independent formal seeds. Err reports the mean and sample standard deviation
over those five formal runs. Because budget selection and formal evaluation use
different seeds, the formal mean MAE may lie above the nominal calibration
target.

As listed in \Cref{tab:iso_accuracy}, \emph{CycleWalker} is the fastest method at all six operating points. It is $10.26\times$--$75.85\times$ faster than tPIRW, and still $1.59\times$--$6.18\times$ faster than the strongest baseline, FasterRW.

\minisection{The tighter the target, the larger the observed advantage}
For a variance-dominated Monte Carlo estimator, the expected stochastic-error
scaling is
\begin{equation}
\mathrm{error}\;\propto\;N^{-1/2},
\label{eq:mc_rate}
\end{equation}
where $N$ is the number of paths. In \Cref{fig:mae_vs_N}, tPIRW and FastRW
closely follow this trend, whereas FasterRW and \emph{CycleWalker} exhibit
method- and case-dependent finite-budget behavior. Consequently, tightening
the target does not multiply every method's budget by an identical factor.

At the six selected operating points, the speedup of \emph{CycleWalker} over
tPIRW increases when the nominal target is tightened from
\SI{0.4}{\kelvin} to \SI{0.2}{\kelvin}. The separation between methods cannot
be attributed solely to per-path variance because the methods also differ in
their priors, residual estimators, covariance treatment, fusion, and kriging.
\Cref{subsec:exp_ablation} isolates several of these mechanisms.

\begin{table}[!t]
\centering
\caption{Matching the deterministic prior's own accuracy.}
\label{tab:prior_iso}
\small
\setlength{\tabcolsep}{5pt}
\begin{tabular}{@{}lcrrr@{}}
\toprule
Method & Err$\pm$Std (K) & $N$ & Time (s) & Speedup \\
\midrule
Periodic prior & $0.0130$ & -- & 1266.86 & 1.00 \\
\textbf{\emph{CycleWalker}} & $\mathbf{0.0126\pm0.0003}$ & \bfseries 5120
  & \bfseries 457.61 & \bfseries 2.77 \\
\bottomrule
\end{tabular}
\end{table}

\Cref{tab:prior_iso} tightens the GPU-HBM-Burst target by another order of
magnitude, to the \SI{0.0130}{\kelvin} MAE reached by the deterministic prior
alone in \SI{1267}{\second}. At $N=5120$, \emph{CycleWalker} obtains
$0.0126\pm0.0003$\,\si{\kelvin} over three seeds in $458\pm12$\,\si{\second}, matching the prior-only reference MAE with a $2.77\times$ runtime reduction.

\subsection{Mechanism Analysis}
\label{subsec:exp_ablation}

Four cumulative configurations are compared on NonAligned and FastCycle over three seeds, each adding one mechanism to the previous one. \textbf{Cutoff} stops every path at the cutoff. \textbf{+Residual} adds whole-path equation-residual estimation. \textbf{+Roulette} adds unbiased Russian roulette. \textbf{+Adaptive} adds pilot-driven sampling and constitutes the full \emph{CycleWalker}. All share an 8-harmonic prior with 32 smoothing sweeps; the first three are fixed at $N=2048$ with cutoff $0.5$, while \textbf{+Adaptive} sets its own budget. In \Cref{tab:ablation}, a bias written $<x$ lies below what three seeds can resolve and is consistent with zero.

\begin{table}[!t]
\centering
\caption{Cumulative ablation of the three mechanisms.}
\label{tab:ablation}
\footnotesize
\setlength{\tabcolsep}{3pt}
\begin{tabular}{@{}llrrrrrr@{}}
\toprule
Case & Cfg. & $N$ & Var. & $|$Bias$|$ & Len. & MAE & Time \\
     &      &     & (K$^2$) & (K) & (steps) & (K) & (s) \\
\midrule
\multirow{4}{*}{NonAligned}
 & Cutoff    & 2048 & 1377.7 & $<$4.0e$-$2 & 3923 & 0.6214 & 17.7 \\
 & +Residual & 2048 & 0.576  & 3.4e$-$2    & 3923 & 0.0342 & 31.7 \\
 & +Roulette & 2048 & 0.704  & $<$9.1e$-$4 & 6585 & 0.0144 & 39.4 \\
 & +Adaptive & 4096 & 0.705  & $<$6.4e$-$4 & 6587 & \bfseries 0.0102 & 55.6 \\
\cmidrule(l){2-8}
\multirow{4}{*}{FastCycle}
 & Cutoff    & 2048 & 1392.4 & $<$4.3e$-$2 & 3923 & 0.6299 & 17.6 \\
 & +Residual & 2048 & 1.849  & 3.0e$-$2    & 3923 & 0.0349 & 27.7 \\
 & +Roulette & 2048 & 2.291  & $<$1.8e$-$3 & 6586 & 0.0260 & 38.4 \\
 & +Adaptive & 8192 & 2.292  & $<$8.8e$-$4 & 6586 & \bfseries 0.0136 & 112.6 \\
\bottomrule
\end{tabular}
\end{table}

\minisection{Residual Estimation}
Going from \textbf{Cutoff} to \textbf{+Residual} in \Cref{tab:ablation}, where
path count, cutoff and mean length are held equal, the per-path variance drops
by $753\times$--$2392\times$ and the MAE by $18\times$. Since only the sampled
quantity differs, we attribute the whole gain to it, and conclude that the
prior need not be accurate: absorbing the dominant component suffices.

\minisection{Unbiased Roulette}
Going from \textbf{+Residual} to \textbf{+Roulette} in \Cref{tab:ablation} at
the same $N=2048$, biases of \SI{3.4e-2}{\kelvin} and \SI{3.0e-2}{\kelvin} fall below what the seeds resolve, paths grow $1.68\times$ longer, and the MAE
improves $2.4\times$ and $1.3\times$. We conclude that roulette removes the truncation bias.

\minisection{Adaptive Sampling}
Going from \textbf{+Roulette} to \textbf{+Adaptive} in \Cref{tab:ablation}, $N$ is raised to 4096 and 8192
for a \SI{0.035}{\kelvin} standard-error target, and the MAE improves
$1.4\times$ and $1.9\times$. Since $N$ is not hand-set here yet the \SI{0.035}{\kelvin} standard-error target is still met, we conclude that adaptive sampling adjusts $N$ automatically, so the user only needs to specify the target.

\begin{table}[!t]
\centering
\caption{Prior sensitivity at a fixed accuracy target.}
\label{tab:prior}
\small
\setlength{\tabcolsep}{3pt}
\begin{tabular*}{\columnwidth}{@{\extracolsep{\fill}}llrrrrr@{}}
\toprule
Method & Prior & $N$ & Var. & MAE & Time & Speedup \\
       &       &     & (K$^2$) & (K) & (s) & \\
\midrule
tPIRW       & none  & 12288 & ---   & 0.186 & 3309.5 & $1.00\times$ \\
\cmidrule(l){2-7}
\multirow{2}{*}{\emph{CycleWalker}}
            & H8+0  & 4096  & 222.6 & 0.186 & 372.4 & $8.9\times$ \\
            & H8+32 & 32    & 1.27  & 0.158 & 43.0  &\textbf{77.0$\times$} \\
\bottomrule
\end{tabular*}
\end{table}

\minisection{Prior Sensitivity}
\Cref{tab:prior} fixes the target at the \SI{0.186}{\kelvin} that tPIRW reaches on
GPU-HBM-Burst, and writes H$L$+$s$ for $L$ harmonics and $s$ sweeps. Dropping the
sweeps inflates the variance $175\times$ and the budget $128\times$, costing
$8.7\times$ runtime, yet leaving an $8.9\times$ margin over tPIRW. We conclude
the prior is indispensable but need not be exact.

\subsection{TSV-Aware Graph Construction}
\label{subsec:exp_tsv}

\Cref{tab:tsv_cost} resolves TSVs under an identical estimator and zero prior, timing five-seed medians at $N=64$ and
accuracy at $N=4096$. The jump rate climbs $22.3\times$ in the TSV cores and
$5.77\times$ globally, so a period costs $5.02\times$ more steps and
$5.25\times$ more GPU time, while their reported MAEs differ by only \SI{0.03}{\kelvin}. We conclude that subgrid homogenization matches the resolved graph's accuracy with half its cells and about one-fifth of its GPU time.

\begin{table}[!t]
\centering
\caption{Cost of resolving TSVs explicitly.}
\label{tab:tsv_cost}
\small
\setlength{\tabcolsep}{5pt}
\begin{tabular}{@{}l@{\hspace{2pt}}rrr@{}}
\toprule
 & Resolved & Subgrid (ours) & Ratio \\
\midrule
TSV-core $\nu_{\mathrm{p99}}$ (\si{\per\second}) & 4.40e5 & 1.97e4 & $22.3\times$ \\
TSV-core $1/\nu_{p99}$ (\si{\micro\second}) & 2.27 & 50.7 & $0.045\times$ \\
Global $\nu_{\mathrm{p50}}$ (\si{\per\second}) & 8471 & 1468 & $5.77\times$ \\
Cells & 175{,}068 & 82{,}420 & $2.12\times$ \\
\midrule
MAE (\si{\kelvin}), $N=4096$ & 0.481 & 0.451 & -- \\
Per-query SE (\si{\kelvin}), $N=4096$ & 0.545 & 0.546 & -- \\
Lookback (periods) & 45.84 & 45.70 & $1.00\times$ \\
\midrule
Steps per period & 1788.9 & \bfseries 356.4 & \textbf{5.02$\times$} \\
GPU time (\si{\second}) & 1.990 & \bfseries 0.379 & \textbf{5.25$\times$} \\
\bottomrule
\end{tabular}
\end{table}
\section{Conclusion}
\label{sec:conclusion}

This paper presented \emph{CycleWalker}, a random-walk solver that evaluates the PSS
of 2.5D/3D ICs only where and when it is queried. Formulating the PSS as an
infinite-history backward walk removes the artificial warm-up of initial-value
transient walks. Sampling only the equation residual of a periodic prior makes
the per-path variance scale with the prior error rather than the temperature.
Unbiased Russian roulette removes the truncation bias, and a pilot-driven budget
meets a prescribed error target without manual tuning. Subgrid TSV homogenization
keeps the jump rate bounded across the multiscale structures of the stack. At
equal accuracy, \emph{CycleWalker} cuts end-to-end solver time by
$10.26\times$--$75.85\times$ against tPIRW, and by $42.62\times$--$75.85\times$
at the tighter \SI{0.2}{\kelvin} target, making it a practical tool for periodic
thermal analysis of 2.5D/3D ICs, especially when only a few locations and phases are queried.

\clearpage
\bibliographystyle{IEEEtran}
\bibliography{ref/Top-sim, ref/bib}

\end{document}